\documentclass[
reprint,
showpacs,
 amsmath,amssymb,
prl
]{revtex4-1}

\usepackage{graphicx}
\usepackage{dcolumn}
\usepackage{bm}
\usepackage{nicefrac}
\usepackage[mathlines]{lineno}
\usepackage{caption}
\usepackage{subcaption}

\begin{document}
\preprint{APS/123-QED}

\title{Nonlinear Optimal Filter Technique For Analyzing Energy Depositions\\ In TES Sensors Driven Into Saturation}

\author{B. Shank}
 \email{bshank@stanford.edu}
\author{J. J. Yen}%
 \author{B. Cabrera}
 \author{J. M. Kreikebaum}
 \author{R. Moffatt}
 \author{P. Redl}
\affiliation{Dept. of Physics, Stanford University, Stanford, CA 94305}

\author{B. A. Young}
\affiliation{Dept. of Physics, Santa Clara University, Santa Clara, CA 95053}

\author{P. L. Brink}
\author{M. Cherry}
\author{A. Tomada}
\affiliation{SLAC National Accelerator Facility, Menlo Park, CA 94025}


\date{\today}

\begin{abstract}
We present a detailed thermal and electrical model of superconducting transition edge sensors (TESs) connected to quasiparticle (qp) traps, such as the W TESs connected to Al qp traps used for CDMS (Cryogenic Dark Matter Search) Ge and Si detectors. We show that this improved model, together with a straightforward time-domain optimal filter, can be used to analyze pulses well into the nonlinear saturation region and reconstruct absorbed energies with optimal energy resolution.
\end{abstract}

\maketitle

\section{Model of Phonon Sensor}

CDMS (Cryogenic Dark Matter Search) relies on superconducting W transition-edge sensors (TESs) connected to Al collector fins to measure energy deposited as hot phonons in Si and Ge substrates by potential dark matter collisions.\cite{CDMS} For a voltage-biased TES, small changes in temperature yield measurable changes in current. Good energy resolution requires small TESs, but finite cross-section to rare particle interactions requires large detectors. To bridge these competing design criteria, CDMS uses fins of superconducting Al coupled to 2-$\mu$m-wide W-TESs. In these detectors, phonons created by an event propagate to the detector surface where some break Cooper pairs in the Al, forming quasiparticles (qp). The qp's diffuse to an Al-W interface where they are trapped in the lower gap W, and heat the W electrons. However, qpÕs that are trapped in the Al in local gap variations do not reach the Al-W interface and their energy is lost. The model described in this paper was used to analyze data from a recent study of the energy collection in CDMS-style W/Al QETs (\textbf{Q}uasi-particle Trap Assited \textbf{E}lectrothermal Feedback \textbf{T}ransition Edge Sensors) by Yen\cite{Yen} where collimated 2.62 keV Cl K$_\alpha$ x-rays were used to study the energy response of square W-TESs (250 $\mu$m on a side) at the ends of 300 nm-thick Al films of different lengths on Si substrates.

In typical voltage-biased operation, a TES is held in its superconducting transition using negative feedback, whereby Joule heating balances the TES power loss to the substrate (defined by $\kappa$ and $n$ below): 
\begin{eqnarray}
P_{Joule} = P_{substrate} \rightarrow
I_{TES}^2 R_{TES} = \kappa T_{TES}^n
\end{eqnarray}
In the case of W below 100 mK, the limiting energy loss mechanism is electron-phonon coupling, with $n$=5.\cite{Saab} Negative feedback also speeds up the return of a perturbed TES to its quiescent state. The characteristic recovery time for electrothermal feedback (ETF) in a TES is:
\begin{equation}
\tau_{\textrm{etf}}=\frac{\tau_0}{1+\nicefrac{\alpha}{n}}
\end{equation}
where $\tau_0$ is the intrinsic thermal time constant \nicefrac{C}{G} and $\alpha \equiv \frac{\partial(log R)}{\partial(log T)} = \frac{T}{R}\frac{\partial R}{\partial T}$ is the unitless steepness parameter for the shape of the resistive transition. For small energy depositions, and near-constant $T_{TES}$, the energy deposited in the TES is simply the decrease in Joule heating integrated over the pulse. In practice, such estimates are systematically low for pulses that span a significant portion of the transition region. Additionally, integrating the pulse results in substantially worse energy resolution than any filter technique where the spectrum of the noise versus that of the pulse is taken into account. For this nonlinear and in principle non-stationary problem, template matching to simulated pulses provides the optimal filter.\cite{Fixsen}

\section{Model Of QET Device}

\begin{figure}[htb]
\begin{center}
\includegraphics[width=8cm]{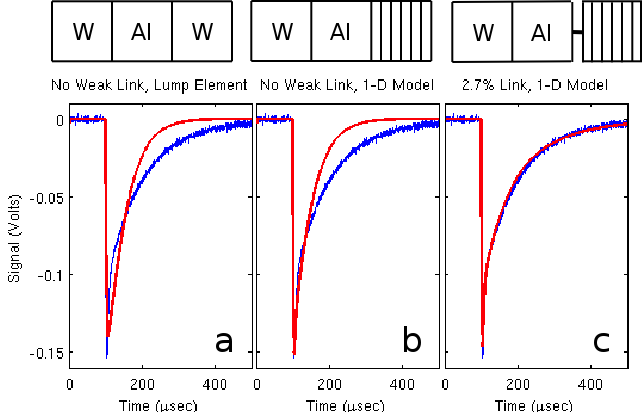}
\end{center}
\captionsetup{justification=raggedright, singlelinecheck=false}
\caption{Simulated pulses (red) and data (blue) for A) a naive lump element model, B) a 1-D model and C) a 1-D model with a weak link at the Al-W interface.}
\label{schematic}
\end{figure}

The earliest model of Al-W qp devices assumed a uniform sheet of current flowing from the Al to the W TES, which then warmed as a single lump element. Results of this model for small energy depositions are shown in Fig. \ref{schematic}a along with data from an actual device. In 2005, Pyle\cite{Pyle} showed that sharp initial spikes observed in real data were a result of the fast but non-instantaneous conduction of heat across the W-TES. In his revised model, the TES was divided into strips along its length and the thermal conduction between strips was found using the measured TES normal-state resistance and the Weidemann-Franz Law. The revised model was better, but it still did not reproduce pulse decays accurately (see Fig.\ref{schematic}b). Further improvements, described here, were made after SEM data\cite{Yen} showed that, due to step-coverage issues, the 40 nm-thick W-TESs in many of our devices were connected to their adjoining 300 nm-thick Al films by W filaments alone ($\sim$2.7\% equivalent coupling for the devices studied here). Such film constrictions increase the local current density and reduce the T$_c$ of the film in that region. This effect creates a small normal region that acts as a heater and allows the TES to lie below the steep part of its transition without going fully superconducting. The resulting reduction in ETF leads to increased pulse decay times. Figure \ref{schematic}c shows that this new model fits our experimental data well.

Because pulse shapes are linked to device temperature, energy reconstructions are potentially sensitive to the heat capacity of the W-TES. Below, we adopt a BCS-like\cite{Kittel} model for the normal and superconducting states:
\begin{eqnarray}
&C_{n}(T) = \gamma T\\
&C_{s}(T) = aT^{-3/2}e^{-\Delta/k_BT} 
\end{eqnarray}
In the normal state, $\gamma = 0.85 meV/mK^2/\mu m^3$ for single-phase, bulk W.\cite{Kittel} In order to achieve a linear energy scale after matching data, we use about half of this $\gamma$ value. The discrepancy is likely due to the polycrystalline properties of our sputtered films. The constant $a$ that sets the scale for $C_{s}$ is computed from Ginzburg-Landau theory while holding the wave number constant through the transition and minimizing the free energy\cite{Goodstein}, yielding $C_s(T_c) = 2.43C_n(T_c)$. To avoid a discontinuity which clearly does not appear in the data, we adopt a two-fluid model. Taking the normal fraction $f_n$ to be some function of the resistance, we have:
\begin{equation}
C_{TES}(T) = f_nC_n(T) + (1-f_n)C_s(T)
\end{equation}
In a uniform-current, large-device approximation, $\it e.g.$ a vortex-induced resistance model\cite{Bardeen}, $f_n = \nicefrac{R_{TES}}{R_n}$ 
More complicated forms for $f_n$ can also be used\cite{Lindeman}. We find that energy reconstructions are relatively insensitive to the shape of $f_n$ as long as other device parameters are fit to data after that choice is made.

\section{Template Matching}

When matching a signal $S_i$ to a series of templates $T_{i,j}$ a standard procedure is to minimize $\chi^2$:
\begin{equation}
\chi_j^2 = \sum_i{\left(\frac{S_i-T_{i,j}}{\sigma_{i,j}}\right)^2}
\end{equation}
where $\sigma_{i,j}$ is the expected rms noise at each template point. In a typical TES system, inherent noise sources include Johnson noise ($V_{rms}=\sqrt{4k_B(\sum_i{R_iT_i})f}$, where $f$ is the inverse of twice the sampling rate) and thermal fluctuations in the link to the thermal bath ($P_{rms}=\sqrt{4k_BT^2gf}$, where $g\equiv \frac{dP}{dT}=n\kappa T^{n-1}$). For our voltage-biased TESs the output is measured as a current, so $I = \nicefrac{V}{R}$.  We construct two independent noise terms:

\begin{eqnarray}
\sigma_{V} &=& \frac{V_{rms}}{R_{TES}}\\
\sigma_{P} = \frac{\partial I}{\partial E}\delta E &=& \frac{\partial I}{\partial T}\frac{\partial T}{\partial E}P_{rms} \Delta t\nonumber\\ = \frac{\partial I}{\partial R}\frac{\partial R}{\partial T}\frac{P_{rms} \Delta t}{C_e} &=& \frac{V}{R}\left(\frac{1}{R}\frac{\partial R}{\partial T}\right)\frac{P_{rms} \Delta t}{C_e}\nonumber\\
\sigma_{P} &=& \frac{I\alpha}{T_e}\frac{P_{rms} \Delta t}{C_e}
\label{eq:deltaI}
\end{eqnarray}
Here $C_e$ is the TES electron heat capacity, $\alpha \equiv \tfrac{T}{R}\tfrac{\partial R}{\partial T}$  and $\Delta t$ is the sampling rate. Although many recent efforts have been made to empirically map out $R(T,I)$ for superconducting transitions, we use here a model\cite{Burney} motivated by Ginzburg-Landau theory.

\begin{equation}
R(T,I) = \frac{R_n}{2}\left(1 + \tanh\left(\frac{T-T_c+(I/A)^{2/3}}{2\ln(2)T_w}\right)\right)
\end{equation}
for a TES with normal resistance $R_n$, critical temperature $T_c$ and 10-90\% transition width $T_w$. The constant A denotes $\frac{I_c}{T_c^{3/2}}$, the strength of the suppression of $T_c$ by non-zero current density in the film.

\section{Non-Stationary Noise}
Defining a variance $\sigma_i^2 = \langle (S_i - T_i)^2 \rangle$ is sufficient if the noise varies so rapidly as to be uncorrelated between measurements. But the possibility of large thermal fluctuations that dissipate on a time-scale $\tau_{ETF} >> \Delta t$ calls for a covariance matrix with its goodness of fit metric:
 \begin{eqnarray}
 \Sigma_{i,j}^2 = \langle (S_i - T_i)(S_j - T_j) \rangle\\
 \chi^2 = (S - T)^T W (S - T)
 \end{eqnarray}
where the weighting matrix $W \propto 1/\sigma^2$ is the inverse of the covariance matrix $\Sigma^2$. In the same way that rms noise from different sources are added in quadrature, covariance matrices from different sources, $\it e.g.$, TES and amplifier, can be added linearly ($\Sigma^2 = \sum_i \Sigma_i^2$). In principle, each element of the simulation could have an independently calculated covariance matrix, but once a simulator with the relevant physics and noise terms is created, it is computationally less costly to make a noiseless template $T_{i,j}$, add a few thousand noisy pulses $S_{i,j,k}$ at each of a comb of energies $E_j$, and calculate the weighting matrices $W$ by Monte-Carlo. The (i,j,k) indexes represent time bin, input energy and pulse number, respectively. Since the energy of real pulses will fall between energies on the comb, we minimize $\chi^2$ by parabolic or third-order fitting to $\chi^2(E)$.

\begin{figure}[htb]
\begin{center}
\includegraphics[width=8.5 cm]{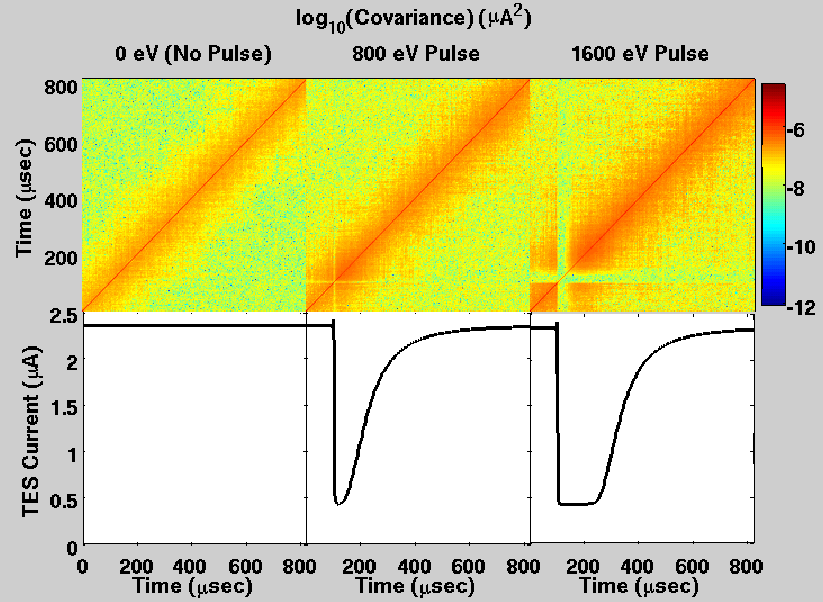}
\end{center}
\captionsetup{justification=raggedright, singlelinecheck=false}
\caption{Covariance matrices (top) for pulses (bottom) of three energies. Color map on log scale. Long-range correlations are suppressed at saturation. A standard $\chi^2$ analysis would use only the diagonal elements.}
\label{covariance}
\end{figure}

Figure \ref{covariance} shows covariance matrices for small, medium and large event energy templates. As expected, the 0 eV template is diagonal (i.e. stationary) with a width of $\sim$100 $\mu$sec $\approx \tau_{ETF}$. As the pulse approaches saturation (middle pane), the diagonal is suppressed, although too little to see in the figure. This is the quantity that would be used in a traditional $\chi^2$ calculation. Most strikingly, during saturation (right pane) the off-diagonal elements are suppressed. Off the transition curve, power fluctuations have negligible coupling to the current readout, so correlations on the scale of $\tau_{ETF}$ are essentially absent.  This feature shows the extent to which non-stationary noise matters for a saturated TES.
\section{Energy Resolution}

\begin{figure}[htb]
\begin{center}
\includegraphics[width=8cm]{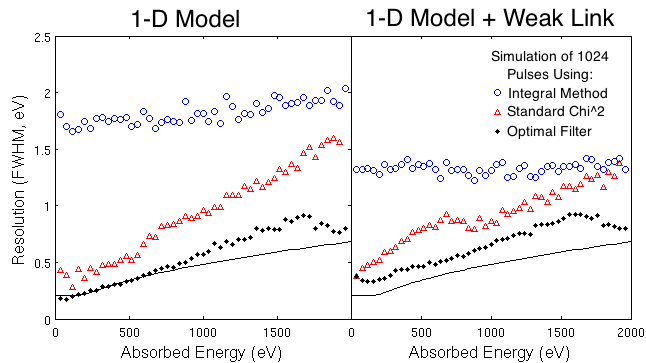}
\end{center}
\captionsetup{justification=raggedright, singlelinecheck=false}
\caption{FWHM energy resolution of covariance $\chi^2$ (black dots), standard $\chi^2$ (red triangles) and integral method (blue circles) reconstructions of 1024 simulated pulses for: (left) ideal link from Al-W qp trap to W-TES, and (right) weak-link model of Fig. \ref{schematic}. The black line shows the calculated theoretical noise limit.}
\label{resolution}
\end{figure}

Models of TES energy resolution are well known\cite{Moseley}. Here we adopt a small-signal model by Irwin\cite{Irwin}, applied specifically to our two regimes of interest:
\begin{equation}
\Delta E_{rms} = \sqrt{\frac{4k_BT_0^2C_0}{\alpha}\sqrt{\frac{n}{2}}} \hspace{20pt} (E << E_{sat})
\label{Erms}
\end{equation}
where $T_0$ and $C_0$ refer to the temperature and heat capacity of the device in its quiescent state. But Eq.\ref{Erms} is only valid for small pulses under the quasi-equilibrium assumption that the device \emph{has} a single temperature at quiescence. At some point, the energy resolution is limited by the ability of ETF to cool the TES. For E $>$ $E_{sat}$, we set $E_{sat}$=$E_{xray}$ in Eq. \ref{Erms}, giving:
\begin{equation}
\Delta E_{rms} = \sqrt{4k_BT_0E\sqrt{\frac{n}{2}}} \hspace{20pt} (E > E_{sat})
\label{ErmsLarge}
\end{equation}
Fig. \ref{resolution} shows the energy resolution achieved by the methods described above when attempting to reconstruct the energy of simulated pulses with added noise. We used a 1-D device model for conduction across the TES and assume no amplifier noise. The black line marks the theoretical best possible resolution. For the integral method, the resolution was scaled-up as if we had adjusted real pulse integrals for a known energy loss computed from the model. For perfect connection (left pane) the model slightly outperforms the theory in the small-pulse limit. This improvement is a reflection of the fact that for these parameters, almost no heat reaches the part of the TES farthest from the Al film, effectively reducing the volume of the W.

To process data through the optimal filter in a reasonable time, both real and simulated pulses are reduced from 4096 to 256 time bins and weighting matrices are 256 bins square. Deviation of the optimal filter performance from theory at high energies is likely due to loss of high frequency information in the down-selection process, which limits our ability to detect the end of the saturation region. In the weak-link model (Fig. \ref{resolution}, right), the increased current density in the link can drive it normal even in the quiescent state. This $\sim$ 0.2 $\Omega$ normal section creates a quiescent temperature gradient across the TES. The excess heat dumped into the TES where it should have the most suppressed Joule heating after an event degrades the ETF. This is especially damaging for small pulse reconstructions where peak shape is important. The extra heat also allows a TES to sit lower in its transition without going fully superconducting. The reduced transition steepness, $\alpha$, reduces the effect of $P_{rms}$ on the integral method (Eq. \ref{eq:deltaI}), although this effect is partially offset by the increased energy loss, particularly at low energies.

\section{Energy Scale}
\begin{figure}[htb]
\begin{center}
\includegraphics[width=8.7cm]{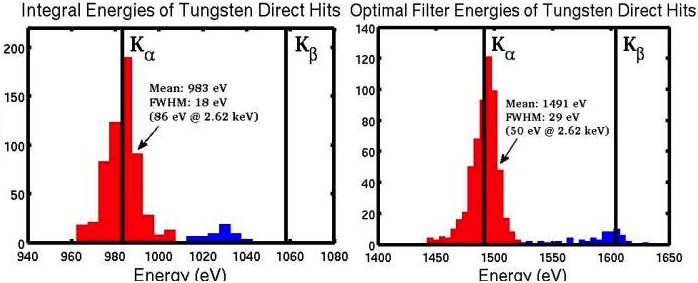}
\end{center}
\captionsetup{justification=raggedright, singlelinecheck=false}
\caption{Reconstructed energy of W-TES direct-hit events using integral (left) and optimal filter (right) methods. FWHM at 2.62 keV scaled to known K$_\alpha$-K$_\beta$ separation.}
\label{fig:hist}
\end{figure}

When an x-ray strikes a metal film, a fraction of the energy gets deposited in the electron system, as modeled in detail by Kozorezov\cite{Kozorezov}. In experiments by Yen, {\it et. al.}\cite{Yen} $\sim$49\% of the energy deposited directly into W-TESs by Cl K$_\alpha$ and K$_\beta$ x-rays was converted to phonons. Moreover, the total event energy, as reconstructed using the optimal filter method presented in this paper, was consistent with Kozorezov's model. Histograms of the data are shown in Fig. \ref{fig:hist}. It is apparent that the optimal filter correctly separates the two x-ray peaks while the integral method does not. We believe the improved resolution seen with the optimal filter in Fig. \ref{fig:hist} is due more to proper removal of environmental noise \emph{a la} Fixsen\cite{Fixsen} than TES physics. Even so, the covariant approach to template matching shown here is a powerful tool that has improved our understanding of TESs.

\section{Conclusion}
We have shown that our new TES weak-link model captures the relevant physics governing TES behavior and produces good fits to observed data. Matching templates from this model to real data using a time-domain optimal filter yields significantly improved energy linearity and event energy reconstructions for real data.

\section{Acknowledgements}
We thank Alexander Kozoresov, Kent Irwin and Saptarshi Chaudhuri for important discussions. This work was supported by the U.S. Department of Energy (DE-FG02-13ER41918 and DE-SC000984) and by the National Science Foundation (PHY-1102842).

\end{document}